\documentclass[runningheads]{llncs}
\usepackage[T1]{fontenc}
\usepackage{todonotes}
\setuptodonotes{inline}
\usepackage{graphicx}

\PassOptionsToPackage{hyphens}{url}\usepackage{hyperref}

\begin{document}
\title{BowelRCNN: Region-based Convolutional Neural Network System for Bowel Sound Auscultation}
%
%
\author{Igor Matynia\orcidID{0009-0000-5406-2336} and Robert Nowak\orcidID{0000-0001-7248-6888}}

%
\authorrunning{I. Matynia, R. Nowak}
%
\institute{Institute of Computer Science, Warsaw University of Technology\\
\url{igor.matynia.stud@pw.edu.pl}; \url{robert.nowak@pw.edu.pl}}
\maketitle              
\begin{abstract}
Sound events representing intestinal activity detection is a diagnostic tool with potential to identify gastrointestinal conditions. This article introduces \textbf{BowelRCNN}, a novel bowel sound detection system that uses audio recording, spectrogram analysys and region-based convolutional neural network (RCNN) architecture. The system was trained and validated on a real recording dataset gathered from 19 patients, comprising 60 minutes of prepared and annotated audio data. BowelRCNN achieved a classification accuracy of 96\% and an F1 score of 71\%. This research highlights the feasibility of using CNN architectures for bowel sound auscultation, achieving results comparable to those of recurrent-convolutional methods.

\end{abstract}

\section{Introduction}

\subsection{Bowel sounds analysis}

Auscultation of sound events representing intestinal activity, called bowel sounds, is a valuable method for assessing intestinal activity. Bowel sounds provide insights into the motor activity of the digestive system. There are four distinct types of bowel sounds: single burst bowel sounds, distinct burst bowel sounds, multiple burst bowel sounds, and continuous random bowel sounds.
\begin{itemize}
	\item Single burst bowel sounds: These are faint and comprise approximately 85\% of all bowel sounds. They occur multiple times per second, typically last 10–40 milliseconds, and have a frequency range of 60 Hz to 2 kHz.
	\item Distinct burst bowel sounds: Louder and more prominent on the spectrogram, these account for about 5–10\% of bowel sounds. Their duration is comparable to single burst sounds, but their frequency can reach up to 3 kHz.
	\item Multiple burst bowel sounds: These represent clusters of single and distinct burst sounds occurring in quick succession. They account for roughly 5\% of all bowel sounds and can last up to 1.5 seconds.
	\item Continuous random bowel sounds: The rarest type, comprising about 1\%, these sounds are irregular and can last for several seconds. They are often associated with audible stomach rumbling.
\end{itemize}

Bowel sounds detection is crucial for monitoring unconscious patients. Moreover such detectors could be used as a noninvasive approach to diagnosing irritable bowel syndrome, a condition that affects 10–15\% of the population. It is especially useful in the case of patients that are unable to communicate their physical symptoms, such as young children \cite{rn:sensors2021review}\cite{rn:sensors2021bowel}.

Despite its potential applications, bowel sound auscultation is not widely adopted in clinical practice. The primary obstacle is the labor-intensive nature of analyzing recordings manually, taking up the valuable time of a medical professional. Those aspects have limited research involving large populations, that could have brought new insights into the intestine activity's relevance to one's health. To address this, an automated system for detecting and analyzing bowel sounds is necessary.
Additionally, reliable detection of bowel sounds is complicated by noise, particularly during extended measurement periods. Common sources of interference include heartbeat, respiratory sounds, clothing friction, and ambient environmental noise.

The objective of this work was to develop a system for the automatic analysis of bowel sounds in audio recordings, enabling the identification of time intervals where these sounds occur. This system is named \textbf{BowelRCNN}. In this research the detection of single-burst bowel sounds will be prioritized. This type of bowel sound appears to be the most relevant for quantitative analysis. Additionally, they are the most common within the dataset and last a relatively short time. They also span a relatively narrow band of the frequency spectrum.

A detection of a bowel sound is considered to be a correctly identified range of time within an audio recording that contains within its boundaries a single identified bowel sound. These detections over the entire length of the recording can be further analyzed to deduce the patients intestine activity.

To facilitate accurate and automated detection of single-burst bowel sounds, machine learning techniques will be applied, specifically convolutional neural networks (CNNs). Spectrogram analysis reveals additional noise below 200 Hz, often attributed to heartbeat and venous hum. The detection system will be developed using Python programming language, as it is widely adopted language for machine learning and signal processing. Spectrograms will be utilized as input data, providing a time-frequency representation of audio signals.

The presented system is a successor to the project carried out by our team \cite{rn:sensors2021bowel}, in which recurrent networks were used.

\section{Methods: BowelRCNN - the bowel sound detection system}

The audio data from an intestinal sound-dedicated contact microphone used in our previous work \cite{rn:sensors2021bowel}
was converted into a single channel WAV, this data is the input to the preprocessing, depicted in section~\ref{sec:preprocessing}, next the detector using artificial neural network, depicted in section~\ref{sec:classification} gives system output.

\subsection{Signal preprocessing}
\label{sec:preprocessing}

\begin{figure}[!htb]
    \centering
    \includegraphics[width=0.9\linewidth]{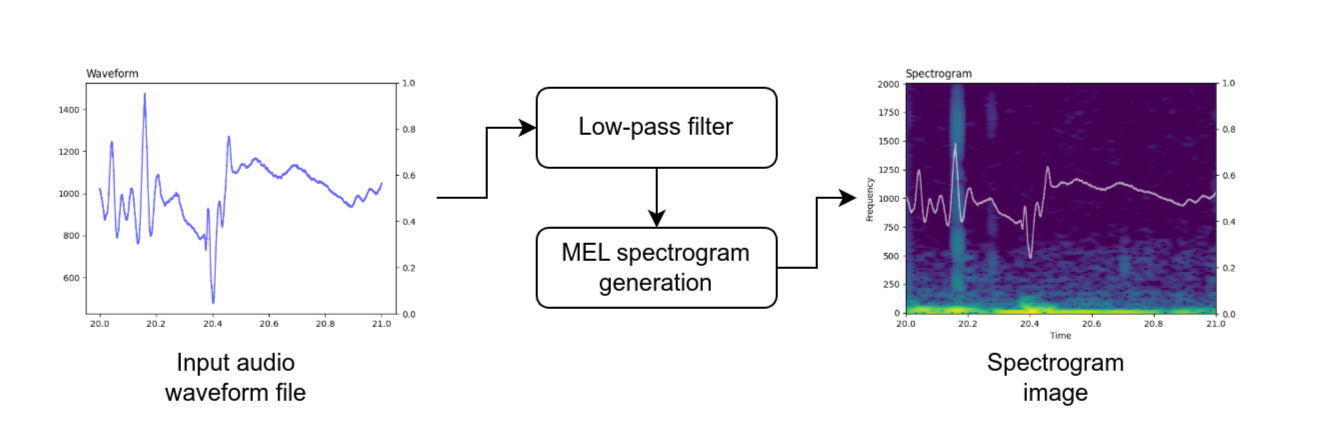}
    \caption{The figure shows the general overview of initial bowel sound recording processing}
    \label{fig:initial-processing}
\end{figure}

An overview of the initial bowel sound recording processing has been shown on Fig.~\ref{fig:initial-processing}.
The initial processing of audio recordings in this research closely follows our previous approach \cite{rn:sensors2021bowel}.
In the current version we have optimized efficiency and scalability, through the use of parallel multi-core computation.

The frequency range of bowel sounds lies between 50 Hz and 2000 Hz, consistent with the characteristics of single burst bowel sounds. To reduce unnecessary data and eliminate high-frequency noise, a low-pass filter with a cutoff frequency of 2000 Hz was applied. This step ensures that frequencies beyond this range, which do not contribute meaningful diagnostic information, are removed before further processing.

To represent the audio data in the time-frequency domain, the recordings were converted into MEL spectrograms using a Hanning (HAN) window. The resulting spectrograms have a resolution of 64 frequency bins and 630 time bins per second of recording. During experimentation an additional spectrogram size has been chosen to investigate its potential benefits.

After spectrogram generation, the resulting spectrograms from all audio files were merged into a single large data structure. Additionally, normalization was applied to the spectrograms within every consecutive 2-second segment of recording.

\subsection{Bowel sound detector}

\begin{figure}[!htb]
    \centering
    \includegraphics[width=0.9\linewidth]{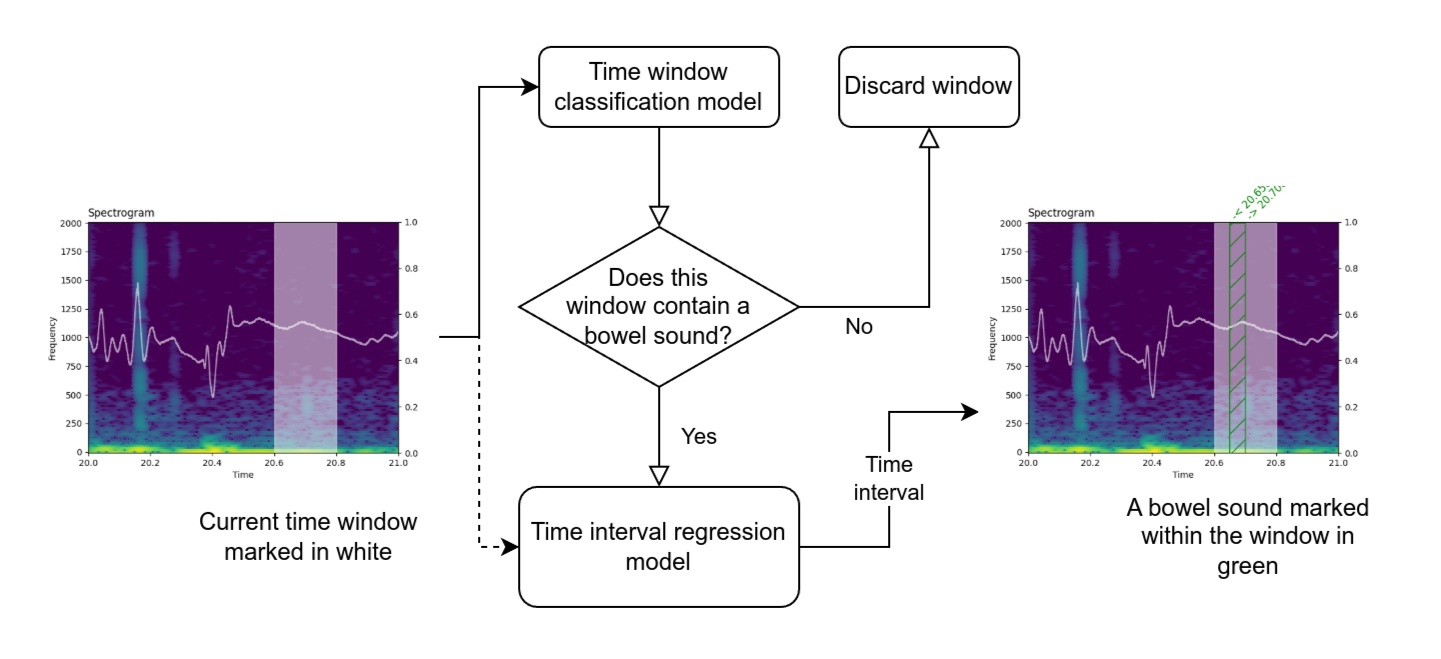}
    \caption{Bowel sound detector. This diagram excludes the initial data processing and predictions aggregation}
    \label{fig:model-overview}
\end{figure}

The detector is divided into two main stages, each utilizing a dedicated convolutional neural network (CNN), as depicted in Fig.~\ref{fig:model-overview}.

The first stage is binary classification of time windows. The model processes the spectrogram divided into short 0.2-second time windows, identifying regions likely to contain bowel sounds. This initial stage acts as a filter, discarding most windows that do not exhibit the desired pattern. The model outputs a binary value indicating the presence or absence of a bowel sound in the analyzed window.

The second stage is precise determination of time intervals. The second model analyzes the windows selected by the first model and accurately determines the time intervals of the bowel sounds within each window by scaling its duration and offset. The model outputs two parameters: a scale factor, which defines the length of the time interval, and an offset, which indicates the sound’s position relative to the center of the window.

\subsection{Training the convolutional neural networks}

The dataset from Kaggle platform \cite{rn:bowel-kaggle-dataset} was used. This dataset consists of audio recordings collected from 19 patients. These recordings were divided into 1605 files, each with a duration of 2 seconds. The entire dataset contains 6413 bowel sounds, from which only those meeting the specified criteria regarding their type and duration were selected. Following the criteria, the remaining amount of sounds available dropped to 6190. The recordings were captured in WAV (waveform) audio file format, with a sampling rate of 44.1 kHz and a 24-bit depth. The annotations for each of the audio files, provided by medical doctors are available and used as ground truth data. The total size of the annotations as well as the audio recordings is 283.82MB. All recordings are anonimised.

The dataset was split into training (70\%), validation (20\%), and test (10\%) sets, ensuring consistent label distribution. Configuration files defined WAV samples and spectrogram parameters for reproducibility. Data processing combined annotations with spectrograms, marking either sound presence or absence. A sampling interface generated time windows with optional Gaussian noise for augmentation, optimizing memory and class balance.

The classification network was trained on a 1:3 ratio of positive (bowel sound) to negative samples using weighted binary cross-entropy, L2 regularization, and dropout to prevent overfitting. Training ran for 250 epochs with a batch size of 256. The regression network, trained on positive samples only, used a modified 1-IOU loss penalizing distance and scale to improve detection in difficult cases. Both networks were trained with a fixed random seed, with all parameters defined via configuration files.

\subsection{Predictions aggregation}

The system predicts bowel sounds by sliding a time window with overlap across the test sample. The classification network first determines if a sound is present; if its probability exceeds a threshold, the regression model refines the time interval. Detected intervals are aggregated by summing overlapping regions, weighted by classification confidence. Those exceeding a vote threshold are reported as positive results.

\section{Results}

Our previous system, depicted in~\cite{rn:sensors2021bowel} was used as baseline method. Both (baseline, and presented approach) were trained and validated on the same data. The networks use the same random seed to ensure reproducibility and share a data augmentation pipeline that includes Gaussian noise to improve robustness against real-world variations. The sampling for the classification network is weighted to select bowel sounds and background noise in a 1:3 ratio. For most experiments, a random number generator seed of 42 is employed.

All subsequent experiments were conducted using prediction parameters set to a threshold of 0.9, a vote fraction of 0.1, and an overlap of 10. The impact of prediction parameters has been analyzed in a separate experiment.

\subsection{Model training parameters optimization}

\begin{table}[!htb]
\caption{Model performance metrics for a given training parameter change. Changes to the baseline model include the learning rate, dropout the max deviation of the Gaussian augmentation}
\label{tab:training-parameters-baseline}
\centering
\begin{tabular}{|c|c|c|c|c|c|c|}
\hline
\textbf{Learning rate} & \textbf{avg\_iou} & \textbf{accuracy} & \textbf{precision} & \textbf{recall} & \textbf{specificity} & \textbf{f1\_score} \\
\hline
0.00005         & 0.524             & 0.966             & \textbf{0.668}    & 0.708         & \textbf{0.981} & 0.687 \\
\textbf{0.0001} & \textbf{0.529}     & \textbf{0.966}    & 0.657             & 0.731         & 0.979         & \textbf{0.692} \\
0.0002          & 0.523              & 0.964             & 0.631             & \textbf{0.754} & 0.976        & 0.687 \\
\hline
\textbf{Dropout} & \textbf{avg\_iou} & \textbf{accuracy} & \textbf{precision} & \textbf{recall} & \textbf{specificity} & \textbf{f1\_score} \\
\hline
\textbf{0.2}    & \textbf{0.532}    & \textbf{0.966}    & 0.650             & \textbf{0.746}    & 0.978             & \textbf{0.695} \\
0.3             & 0.529             & 0.966             & \textbf{0.657}    & 0.731             & \textbf{0.979}    & 0.692 \\
0.5             & 0.526             & 0.965             & 0.648             & 0.737             & 0.978             & 0.690 \\
\hline
\textbf{Gauss std max} & \textbf{avg\_iou} & \textbf{accuracy} & \textbf{precision} & \textbf{recall} & \textbf{specificity} & \textbf{f1\_score} \\
\hline
\textbf{No augment}       & \textbf{0.542}    & \textbf{0.967}    & \textbf{0.671}    & 0.738             & \textbf{0.980} & \textbf{0.703} \\
0.15                        & 0.540             & 0.967             & 0.667             & \textbf{0.739}    & 0.980 & 0.701 \\
0.3                         & 0.529             & 0.966             & 0.657             & 0.731             & 0.979 & 0.692 \\
0.6                         & 0.523             & 0.966             & 0.656             & 0.721             & 0.979 & 0.687 \\
\hline
\end{tabular}
\end{table}

This experiment assessed the impact of learning rate, dropout, and Gaussian augmentation deviation on performance metrics (Table \ref{tab:training-parameters-baseline}), with IoU and F1 as key indicators.

A learning rate of 0.0001 was optimal, balancing precision and recall (F1: 0.692, IoU: 0.529). A lower rate (0.00005) slightly improved precision (0.668) but reduced F1. A dropout rate of 0.2 achieved the best results (IoU: 0.532, F1: 0.695), while higher rates (0.3, 0.5) marginally lowered performance. The highest metrics (IoU: 0.542, F1: 0.703) were observed without Gaussian augmentation, but results with 0.15 std deviation were nearly identical, making it the final choice for added robustness.

\subsection{Tweaked network architecture}

Table \ref{tab:baseline-architectures} presents experiments assessing different CNN architectures while maintaining the baseline spectrogram size. Metrics included IoU, accuracy, precision, recall, specificity, and F1 score.

\begin{table}[!htb]
\caption{Performance Metrics for baseline spectrogram size for different CNN architectures.}
\label{tab:baseline-architectures}
\centering
\begin{tabular}{|l|c|c|c|c|c|c|}
\hline
\textbf{Model id} & \textbf{avg\_iou} & \textbf{accuracy} & \textbf{precision} & \textbf{recall} & \textbf{specificity} & \textbf{f1\_score} \\
\hline
\textbf{Baseline}       & \textbf{0.529} & 0.966        & 0.657         & \textbf{0.731} & 0.979        & \textbf{0.692} \\
Bigger network          & 0.528         & \textbf{0.968} & \textbf{0.689} & 0.694       & \textbf{0.983} & 0.691 \\
Smaller network         & 0.494         & 0.965         & 0.673         & 0.651         & 0.983         & 0.662 \\
Increased CNN layers    & 0.522         & 0.967         & 0.678         & 0.693         & 0.982         & 0.686 \\
MSE loss                & 0.466         & 0.957         & 0.566         & 0.727         & 0.969         & 0.636 \\
\hline
\end{tabular}
\end{table}

The baseline model performed best overall (IoU: 0.529, recall: 0.731, F1: 0.692), offering strong results with minimal parameters. The larger model (740K+ parameters) added filters, an extra CNN layer, and larger fully connected layers, achieving the highest accuracy (0.968) and precision (0.689) but with only marginal gains. The smaller model (140K–313K parameters) underperformed, indicating insufficient capacity. The deeper model (188K parameters, two extra CNN layers) had good metrics (F1: 0.686) but did not surpass the baseline, suggesting diminishing returns. The MSE loss model, replacing weighted IoU with MSE, performed the worst (IoU: 0.466, F1: 0.636), emphasizing the importance of task-specific loss functions.

The baseline model was selected for its strong performance, simplicity, and efficiency, as larger models offered only minor improvements at the cost of added complexity.

\subsection{Increased spectrogram size experimentation}

\begin{table}[!htb]
\caption{Model performance metrics for an increased spectrogram size, for different CNN architectures}
\label{tab:training-big-spec}
\centering
\begin{tabular}{|l|c|c|c|c|c|c|}
\hline
\textbf{Model id} & \textbf{avg\_iou} & \textbf{accuracy} & \textbf{precision} & \textbf{recall} & \textbf{specificity} & \textbf{f1\_score} \\
\hline
Baseline            & 0.175             & 0.922         & 0.281         & 0.318         & 0.955     & 0.298 \\
Bigger network      & 0.184             & 0.922         & 0.290         & 0.334         & 0.955     & 0.310 \\
Smaller network     & 0.174             & \textbf{0.925} & \textbf{0.292} & 0.300       & \textbf{0.960}     & 0.296 \\
MSE loss            & \textbf{0.186}    & 0.918         & 0.279         & \textbf{0.357} & 0.949    & \textbf{0.313} \\
No augmentation     & 0,179             & 0,921             & 0,282     & 0,328           & 0,954                & 0,303              \\
Unmodified IOU      & 0,174             & 0,922             & 0,281     & 0,312           & 0,956               & 0,296            \\ 
\hline
\end{tabular}
\end{table}

The experiment summarized in Table \ref{tab:training-big-spec} evaluates the effect of increasing the spectrogram size to 315 by 126 pixels on model performance across different CNN architectures. The configurations mentioned introduce the same changes as presented in table \ref{tab:baseline-architectures}. Increasing the spectrogram size severely degraded performance across all metrics, indicating reduced generalization capacity of the models. For instance, the highest F1 score (0.313) was achieved by the "MSE loss" configuration, alongside the best average iou (0.186). However, even this performance is notably lower than previous experiments with smaller spectrogram sizes. Similarly, the "Smaller network" achieved the highest accuracy (0.925) and precision (0.292), but overall metrics remained suboptimal.
The architectures used in these experiments remained consistent with those from the previous set. However, the total parameter count increased fourfold due to the spectrogram size doubling in each dimension.

\subsection{Prediction parameters optimization}

A comprehensive experiment tested 240 configurations to optimize prediction parameters for the baseline model. Predictions were averaged over five models trained with seeds 42–46. Three threshold values (0.9, 0.75, 0.5), four overlap values (1, 5, 10, 25), and four vote fractions (0.05, 0.1, 0.2, 0.4) were evaluated for their impact on IoU, F1, accuracy, precision, recall, and specificity.

\begin{figure}[!htb]
    \centering
    \includegraphics[width=0.75\linewidth]{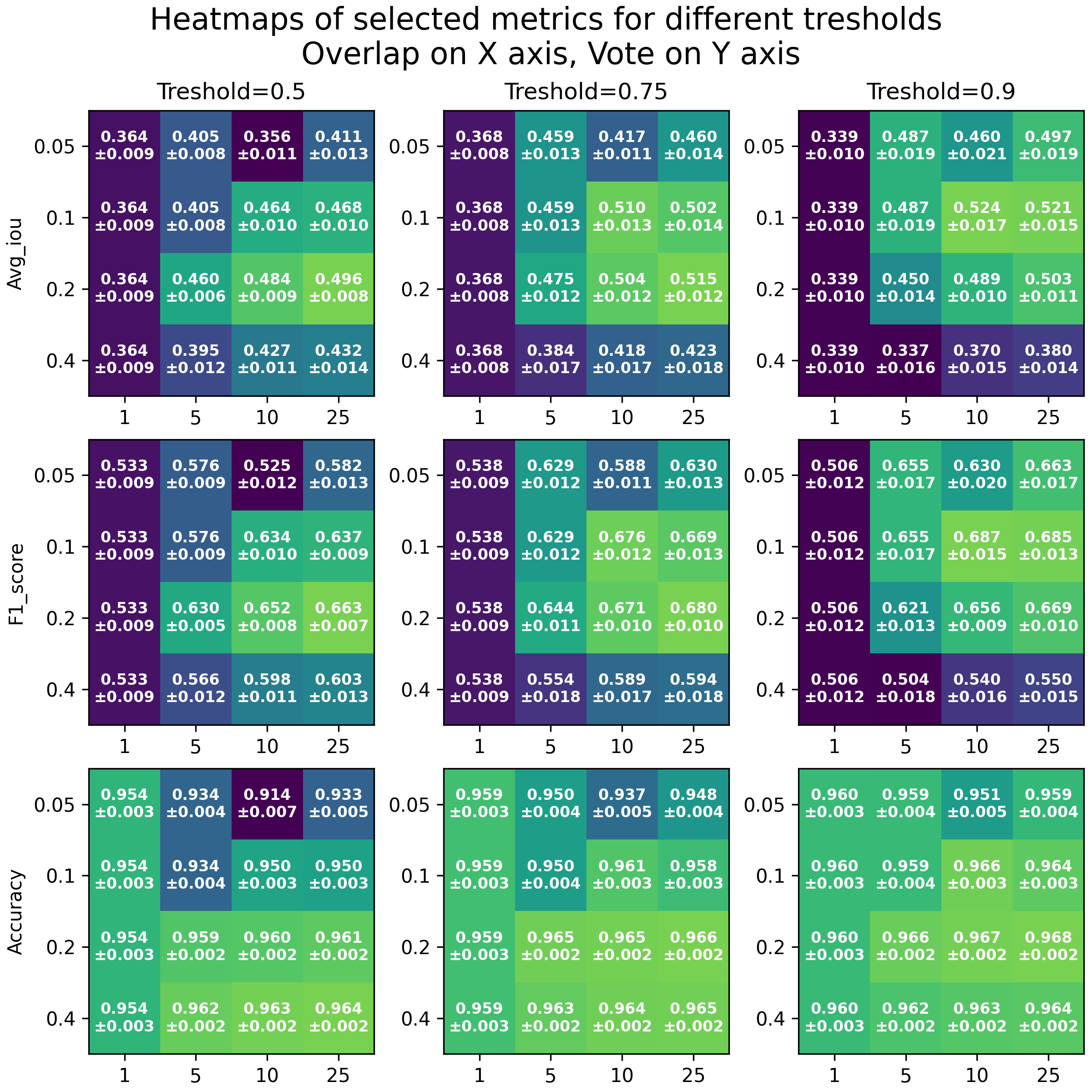}

    \caption{The heatmap shows the values of selected metrics as well as selected detection threshold values on the horizontal axis. The brighter the color of the cell, the higher the value. The values have been averaged over 5 models trained with different random number generator seeds.}
    \label{fig:inference-parameter-all}
\end{figure}

Figures \ref{fig:inference-parameter-all} presents a heatmap summarizing parameter effects on these metrics. Results showed minimal variance across seeds, confirming consistency. The first heatmap columns were identical due to an overlap of 1, making this setting vulnerable to false positives.

A higher vote fraction improved specificity and precision but reduced recall by filtering low-confidence yet correct predictions. Overlap had the strongest impact on IoU and F1, peaking at 10. Increasing the threshold improved precision at the cost of recall. The optimal parameters were identified as threshold = 0.9, vote fraction = 0.1, and overlap = 10.

\subsection{Overall model comparison results}

The final comparison of model performance, presented in Table \ref{tab:comparison}, evaluates the newly developed BowelRCNN, the existing CRNN model (both locally trained from source and as reported in the original work), and two meta-algorithm variations. These results have been collected on the same test dataset, different from the original work.

\begin{table}[!htb]
\caption{Final comparison between best BowelRCNN, existing CRNN\cite{rn:sensors2021bowel} (data gathered on the same test set locally as well as metrics from the original work) and meta-algorithm results}
\label{tab:comparison}
\centering
\begin{tabular}{|l|c|c|c|c|c|c|}
\hline
\textbf{Model id} & \textbf{avg\_iou} & \textbf{accuracy} & \textbf{precision} & \textbf{recall} & \textbf{specificity} & \textbf{f1\_score} \\
\hline
Best BowelRCNN  & 0.551         & 0.968         & 0.682             & 0.742             & 0.981             & 0.711 \\
Meta-intersect  & 0.543         & \textbf{0.974} & \textbf{0.872}   & 0.590             & \textbf{0.995}    & 0.704 \\
Meta-sum        & \textbf{0.570} & 0.967        & 0.646             & \textbf{0.828}    & 0.975             & \textbf{0.726} \\
CRNN local      & 0.566         & 0.973 &        0.777              & 0.676             & 0.989             & 0.723 \\
\hline
CRNN original\cite{rn:sensors2021bowel}& & 0.981 &        0.898              & 0.888             & 0.990             & 0.893 \\
\hline
\end{tabular}
\end{table}

The Best BowelRCNN consisted of: 
\begin{itemize}
    \item The pattern model with 3 convolutional layers with a total of 1,253,170 parameters, using filter sizes of 8, 16, and 16, followed by 2 linear layers with 512 neurons each.
    \item The classification model includes 3 convolutional layers with filter sizes of 8, 16, and 16, and 2 linear layers with 256 neurons each, totaling 661,873 parameters.
\end{itemize}
Both models are trained with a learning rate of 0.0002 and utilize data augmentation via Gaussian noise with the standard deviation of 0.15.

\section{Summary}

This article presents a novel system for automatic bowel sound analysis using convolutional networks. The system operates in two stages: classifying time windows to identify potential bowel sound regions and precisely determining their time intervals. Scripts were developed in Python, for data preparation, model training, prediction generation, and experiment execution. The model achieves competitive results in key metrics (mean IOU, F1, precision, sensitivity) at reduced spectrogram resolution.

Future improvements could include refining the aggregation step by using Gaussian distributions for smoother confidence representation and exploring advanced meta-algorithms for combining model predictions. Expanding the dataset and incorporating diverse data augmentations would enhance robustness. Semi-supervised learning could address limited labeled data, while integrating recurrent or transformer-based architectures might improve context-awareness and temporal dependency modeling.

To increase accessibility, a web-based interface could be developed, enabling non-technical users to interact with the system. Additionally, with its short inference time, the system could be adapted for real-time monitoring, although significant codebase modifications would be required. These advancements would further improve the system's accuracy, usability, and applicability.

\subsubsection*{Author contributions}

R.N. identified the problem, I.M. designed the approach, downloaded the data, implemented the software, performed numerical experiments, I.M and R.N. prepared the draft. All authors have read and agreed to the published version of the manuscript.

\subsubsection*{Funding}
A statutory Research Grant from the Institute of Computer Science, Warsaw University of Technology, supports this work.

\subsubsection*{Software availability}

BowelRCNN is available on the \url{https://github.com/IMatynia/bowelrcnn} repository under the MIT license.

\subsubsection*{The authors declare no conflict of interest.}


\end{document}